# An Alternative Treatment of the Foucault Pendulum

W. Zimmermann Jr.
School of Physics and Astronomy
University of Minnesota
Minneapolis, Minnesota, U.S.A.
Email: zimme004@umn.edu

## Abstract

A treatment is given of the precession of a Foucault pendulum by means of two successive rotational transformations of coordinate system. The simplicity and accuracy of this approach is emphasized.

The Foucault pendulum, a simple pendulum suspended from a fixed point at the surface of the earth and free to swing isotropically in any direction of the compass, is a physical system of continual interest, both for its historical and present-day demonstration of the rotation of the earth and for its role in illustrating the dynamics associated with a rotating coordinate system.[LF51] The pendulum is discussed in many textbooks dealing with classical mechanics, for example references [BO],[FC],[FW],[G],[GPS],[LL],[MT],[S],[SG],[T] as well as in many articles in this journal, for example references [BB07],[C81],[C95],[HMM87],[JM10],[PP08],[SBPC10],[SD70]. Attention is also called to the web articles [H] and [W].

Although there are various ways of treating the motion of the pendulum, the most straight-forward approach, perhaps, is simply to solve the equations of motion that result from a transformation of coordinates to a non-inertial system at rest on the earth's surface at the location of the pendulum. In addition to the force of gravity and the tension in the "string" acting on the "bob", these equations involve the fictitious Coriolis and centrifugal forces. It is usual to work in the small-amplitude limit, so that the locally vertical motion of the bob can be ignored and the problem becomes the same as that of a perturbed two-dimensional isotropic simple harmonic oscillator. It also turns out that to a very good approximation the centrifugal force can be ignored, aside from the slight deviation of the rest position of the pendulum from the local vertical, leaving only the Coriolis force as a disturbance.

The purpose of the present note is to call attention to a simple alternative to the above approach. In brief, this alternative is to follow the first transformation of coordinate system to a system at rest on the earth's surface by a second transformation from that system to a system in which the locally horizontal component of the Coriolis force vanishes. This is a system that is rotating about the local vertical $\hat{z}$-axis with angular velocity

$$\boldsymbol{\omega}_r = -\omega_e \cos\theta \, \hat{\mathbf{z}}_r \quad , \tag{1}$$

where $\omega_e$ is the angular velocity of the earth's rotation and $\theta$ is the colatitude. The direction of this second rotation is clockwise in the northern hemisphere and counter-clockwise in the southern, as seen looking toward the center of the earth. The sum of the two angular velocity vectors then has no vertical component at the location of the pendulum. An approach very similar to this alternative was presented in the excellent classical mechanics text by Symon, in particular.[S] Figures 1 and 2 show the elements of this approach.

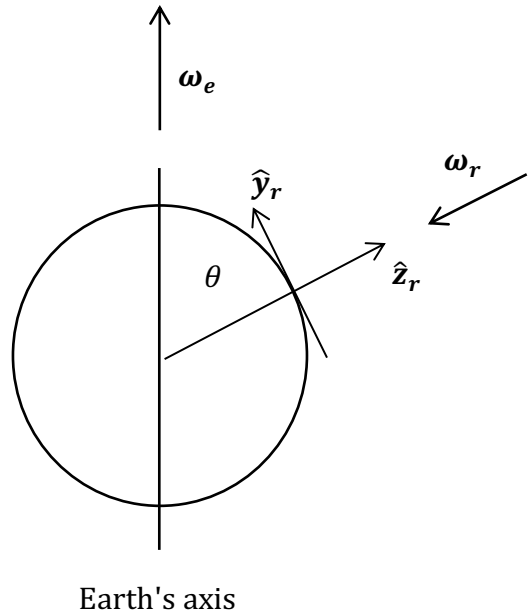

Earth's axis

Fig. 1. Shown here, in section, is the Earth, rotating with angular velocity $\boldsymbol{\omega}_e$ about the polar axis. Attached to it at a local origin at colatitude $\theta$ and rotating with it is the $x_r, y_r, z_r$ coordinate system, with $x_r$ pointing locally east, $y_r$ pointing north, and $z_r$ pointing upward. The vector $\boldsymbol{\omega}_r$ is an angular velocity that when added to $\boldsymbol{\omega}_e$ yields a vector that has only a $y_r$-component and no $z_r$-component.

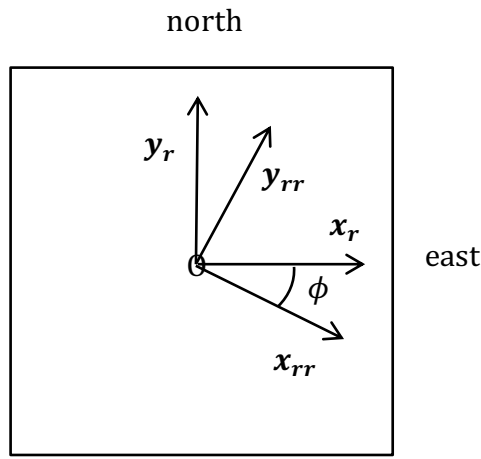

Fig.2. Shown here is the local horizontal $x_r, y_r$ plane fixed to the earth's surface at O and rotating with it, viewed from above. Also shown are the $x_{rr}$ and $y_{rr}$ axes of a coordinate system that rotates relative to the $x_r, y_r$ system about a common z-axis with angular velocity $d\phi/dt = -\omega_r = \omega_e \cos\theta$.

In what follows, it is useful to keep in mind some representative numerical values. The angular velocity $\omega_e$ equals $7.44 \times 10^{-5}\, rad/s$ ($15.35\,°/hr$). For a pendulum whose suspension has a length $l$ of $30\,m$, representative of some of the longer pendulums constructed, the angular velocity $\omega_p = \sqrt{l/g}$ would be $0.57\, rad/s$ and the period of the oscillation $T_p = 2\pi/\omega_p$ would be $11.0\,s$. Assuming the maximum deflection of the bob from equilibrium $a$ to be a modest $0.90\,m$, the ratio $\alpha$ of this deflection to the length would be $0.030$.

In vector form, the full equation needed to convert Newton's second law from an inertial system to a system rotating with angular velocity $\boldsymbol{\omega}$ is one that expresses the second derivative of the position $\boldsymbol{r_a}$ of a mass with respect to time in an inertial system in terms of a relationship involving the first and second derivatives of the position with respect to time in the rotating system

$$d^2\boldsymbol{r_a}/dt^2 = (d^2\boldsymbol{r_a}/dt^2)_r + 2\boldsymbol{\omega} \times (d\boldsymbol{r_a}/dt)_r + \boldsymbol{\omega} \times (\boldsymbol{\omega} \times \boldsymbol{r_a}) + (d\boldsymbol{\omega}/dt) \times \boldsymbol{r_a}. \tag{2}$$

Here $(\,)_r$ denotes evaluation in the rotating system. The second term on the right is called the Coriolis term and the third is the centrifugal term. The last term, in which the derivative is the same evaluated in the rotating system as in the inertial system, will not be needed for the earth's rotation, but we include it here for completeness and later use. It is sometimes called the Euler term.[H],[WP] Note that in this equation, it is assumed that $\boldsymbol{r_a}$ extends out from some point on the axis of rotation.

As we apply the formula above to the Foucault pendulum, let's replace $\boldsymbol{r_a}$ with the sum $\boldsymbol{R_a} + \boldsymbol{r}$, where $\boldsymbol{R_a}$ is a vector from some point on the earth's axis to an origin for the pendulum on the earth's surface and $\boldsymbol{r}$ is now the location of the pendulum bob relative to that origin. Due to the linearity of the equation in terms of the $\boldsymbol{r_a}$ and its derivatives, we now have two independent equations of the same form as that above, one for $\boldsymbol{R_a}$ and one for the new, relative $\boldsymbol{r}$. (This operation is equivalent to a parallel displacement of the axis of rotation.) Since $\boldsymbol{R_a}$ is fixed in the rotating system and $\boldsymbol{\omega}$ is constant, its equation is particularly simple,

$$d^2\boldsymbol{R_a}/dt^2 = \boldsymbol{\omega} \times (\boldsymbol{\omega} \times \boldsymbol{R_a}) = -\omega_e^2 \boldsymbol{R_{a\perp}}, \tag{3}$$

where $\boldsymbol{R_{a\perp}}$ is the component of $\boldsymbol{R_a}$ perpendicular to the axis of rotation. The effect of this relation is to add a small effective force to gravity acting on any mass $m$ at the local origin, equal to $m\omega_e^2 \boldsymbol{R_{a\perp}}$. This effect causes a small shift in the equilibrium position of the pendulum bob in a direction away from the earth's axis, but since this same effect also influences the figure of the surface of the earth, it seems natural to let the equilibrium position of the pendulum define the vertical axis of the local coordinate system. This allows us to ignore this effect from now on, except for a small deviation of the effective local latitude from the geometrical one and an effective gravitational acceleration $g_r$ that is slightly smaller then $g$ due to the earth's gravitational attraction alone. At a latitude of $45°$, the strength of the centrifugal acceleration is only $2.5 \times 10^{-3}$ times the gravitational acceleration $g$ and the angular deviation of our effective vertical axis from a line through the local origin from the center of the earth is only 0.10°. (These effects are too small to appear in Fig.1.)

With the approximations and considerations mentioned above, the equation governing the motion of the pendulum bob in the local horizontal $x_r, y_r$ plane shown in Fig. 2 becomes

$$m(d^2\mathbf{r}/dt^2)_r \cong -(mg_r/l)\mathbf{r} - 2m\,\omega_e\cos\theta\,\hat{\mathbf{z}}_r \times (d\mathbf{r}/dt)_r, \tag{4}$$

where $m$ is the mass of the bob. This is the equation for an isotropic two-dimensional simple harmonic oscillator with an effective perturbing Coriolis force acting sidewise to the velocity, tending to deflect the motion to the right in the northern hemisphere and to the left in the southern. A usual treatment of the Foucault pendulum is then to solve the two coupled equations for $x_r$-axis and $y_r$-axis motion that result from this vector equation of motion.[FC],[FW],[G],[LL],[MT],[SG],[T],[SD70],[H],[W].

In our alternative approach we cause the Coriolis perturbation to all but disappear by a further appropriate rotational transformation to a second coordinate system, one which cancels the local vertical component of the earth's rotation.[FC],[S] This is the system described above and shown by the $x_{rr}$ and $y_{rr}$ axes in Fig. 2, rotating about the local vertical with angular velocity $\boldsymbol{\omega}_r = -\omega_e\cos\theta\,\hat{\mathbf{z}}_r$. This second transformation involves applying once again the same relation written above as Eq.(2) but now in the first rotating system and can be written

$$(d^2\mathbf{r}/dt^2)_r = (d^2\mathbf{r}/dt^2)_{rr} + 2\boldsymbol{\omega}_r \times (d\mathbf{r}/dt)_{rr} + \boldsymbol{\omega}_r \times (\boldsymbol{\omega}_r \times \mathbf{r}) + (d\boldsymbol{\omega}_r/dt)_r \times \mathbf{r}, \tag{5}$$

where now $()_{rr}$ denotes evaluation in the second system. Here, again, the Euler term, which can be evaluated either in the $r$ or the $rr$ system, is zero. When this second transformation is applied to the complete equation of motion resulting from the first transformation, the outcome within the same approximations applied earlier for motion in the local horizontal plane is just

$$m(d^2\mathbf{r}/dt^2)_{rr} \cong -(mg_r/l)\mathbf{r}, \tag{6}$$

the equation of motion of an unperturbed isotropic two-dimensional simple harmonic oscillator with angular frequency $\omega_p = \sqrt{g_r/l}$. Thus we can conclude immediately that the resulting motion of the bob relative to the earth's surface is in general the elliptical motion of the isotropic oscillator superimposed on the slow rotation given by $\boldsymbol{\omega}_r$. The same result is, of course, reached by the usual treatment. The simplest solution is linear motion of the bob in the second system. For this solution, it can be shown, and this demonstration is given in a supplement to this article, that the neglected terms are all quite small in relation to the dominant terms included in Eqs. (4) and (6), involving various factors of $\alpha$ and $\omega_e/\omega_p$. Furthermore, for linear motion in the $rr$ system, none of these terms, small as they are, makes any direct contribution to the rate of precession, which is thus given quite accurately (in principle) by Eq. (1).

It is important to recognize, however, that for elliptical solutions of the pendulum motion in the $rr$ system, the small departure of the pendulum motion from simple harmonic motion neglected above can cause an appreciable precession of the solution that adds to or subtracts from the precession due to the earth's rotation seen in the $r$ system. [FC],[SG],[OL78],[OL81],[C81] For small oscillations of the pendulum, the angular velocity of this "natural" precession is given to lowest order by the expression

$$\Omega = (3/8)(ab/l^2)\omega_p , \qquad (7)$$

where $a$ is now the semi-major axis of the ellipse and $b$ the semi-minor axis in the horizontal plane.[OL78],[OL81] To the same order of approximation, the square of the angular frequency of oscillation is lowered to

$$\omega^2 = \omega_p^2[1 - (a^2 + b^2)/8l^2] . \qquad (8)$$

Use of the representative values given earlier yields the result that for $\Omega$ to equal the rate of precession due to the Coriolis force at a latitude of 45°, $b$ must equal 0.25 m, so that to minimize the influence of natural precession, $b$ must be kept much less than this value. If the pendulum motion is planar in the $rr$ system, this precession is absent. However, it is often the case that a pendulum is first released from rest relative to the earth's surface. (Often by burning a thread that holds the pendulum deflected at rest.) The solution in this case then requires a small elliptical departure from linear motion of the bob to be superimposed on the steady precession in order to satisfy the initial conditions. In terms of the representative values used above, the resulting semi-minor axis $b$ would be only $8.3 \times 10^{-5}\, m$, yielding a very small addition to the precession rate due to the Coriolis force. In practice, the precession is often influenced more strongly by small anisotropies in the suspension.

The illustrative parameters and idealizations that we have adopted paint a rather favorable picture. A great deal of ingenuity has been employed in overcoming some of the practical problems that arise, particularly in shorter pendulums, and pendulums that incorporate measures to overcome damping of the motion and keep them operating in steady-state. See e.g. references [C81],[C95],[PP08],[SBPC10],[SD70].

The reader may well wish to stop at this point, the main message of this article having been delivered. However, as a coda to this treatment, it is interesting to consider more completely the successive application of two rotational transformations. Let's designate the first rotation $\boldsymbol{\omega}_1$, the second $\boldsymbol{\omega}_2$, and derivatives evaluated in the final system $()_{rr}$. The full result of this succession of transformations is

$$\frac{d^2\boldsymbol{r}}{dt^2} = \left(\frac{d^2\boldsymbol{r}}{dt^2}\right)_{rr} + 2(\boldsymbol{\omega}_1 + \boldsymbol{\omega}_2) \times \left(\frac{d\boldsymbol{r}}{dt}\right)_{rr} + \boldsymbol{\omega}_1 \times (\boldsymbol{\omega}_1 \times \boldsymbol{r}) + \boldsymbol{\omega}_2 \times (\boldsymbol{\omega}_2 \times \boldsymbol{r}) + 2\boldsymbol{\omega}_1 \times (\boldsymbol{\omega}_2 \times \boldsymbol{r})$$

$$+[(d\boldsymbol{\omega}_1/dt)_1 + (d\boldsymbol{\omega}_2/dt)_2] \times \boldsymbol{r}. \tag{9}$$

The designation $(d\boldsymbol{\omega}_1/dt)_1$ signifies the time derivative of $\boldsymbol{\omega}_1$ that applies in the first transformation and similarly for $(d\boldsymbol{\omega}_2/dt)_2$ in the second. We see from the presence of the next to last term that this expression is not quite symmetrical in $\boldsymbol{\omega}_1$ and $\boldsymbol{\omega}_2$. The result depends on the order in which the successive rotations are applied!

In order to test our understanding of the situation, let's consider a single transformation involving $\boldsymbol{\omega} = \boldsymbol{\omega}_1 + \boldsymbol{\omega}_2$. The result is

$$\frac{d^2\boldsymbol{r}}{dt^2} = \left(\frac{d^2\boldsymbol{r}}{dt^2}\right)_{rr} + 2(\boldsymbol{\omega}_1 + \boldsymbol{\omega}_2) \times \left(\frac{d\boldsymbol{r}}{dt}\right)_{rr} + (\boldsymbol{\omega}_1 + \boldsymbol{\omega}_2) \times [(\boldsymbol{\omega}_1 + \boldsymbol{\omega}_2) \times \boldsymbol{r}] + [d(\boldsymbol{\omega}_1 + \boldsymbol{\omega}_2)/dt] \times \boldsymbol{r}. \tag{10}$$

This expression appears symmetrical in $\boldsymbol{\omega}_1$ and $\boldsymbol{\omega}_2$, and we might be tempted to think that it provides a route to successive transformations independent of the order in which they are applied. But in such cases we need to think carefully about the last term. As above, it can be evaluated either in the original or the final system with equal results. If we evaluate it in the original system, $d\boldsymbol{\omega}_1/dt$ here equals $(d\boldsymbol{\omega}_1/dt)_1$ above, while $d\boldsymbol{\omega}_2/dt$ equals $(d\boldsymbol{\omega}_2/dt)_2 + \boldsymbol{\omega}_1 \times \boldsymbol{\omega}_2$. Thus $d(\boldsymbol{\omega}_1 + \boldsymbol{\omega}_2)/dt$ here equals $(d\boldsymbol{\omega}_1/dt)_1 + (d\boldsymbol{\omega}_2/dt)_2 + \boldsymbol{\omega}_1 \times \boldsymbol{\omega}_2$. If we evaluated $d(\boldsymbol{\omega}_1 + \boldsymbol{\omega}_2)/dt$ in the final system, we would obtain the same result. So here again, the order in which the transformations are applied matters. Reassuringly, the final expression for $d^2\boldsymbol{r}/dt^2$ that is obtained from Eq.(10) is the same as that obtained above from Eq.(9).

To go back briefly to our particular succession of rotational transformations, where $\boldsymbol{\omega}_1 = \boldsymbol{\omega}_e$ and $\boldsymbol{\omega}_2 = \boldsymbol{\omega}_r$, both $(d\boldsymbol{\omega}_1/dt)_1$ and $(d\boldsymbol{\omega}_2/dt)_2$ are zero whereas $d(\boldsymbol{\omega}_1 + \boldsymbol{\omega}_2)/dt$ is not; from the initial inertial system $\boldsymbol{\omega}_2$ appears to be time-dependent, while from the final $rr$ system $\boldsymbol{\omega}_1$ appears to be time-dependent.

Supplement to "An Alternative Treatment of the Foucault Pendulum"

The Neglected Terms

In this supplement, we apply the full Eq.(9) or (10) to the case of the Foucault pendulum, including contributions from the higher-order terms that can usually be neglected. Our ultimate goal will be to look to see whether any of these terms acts to alter the rate of precession, small as they are. In working out the contributions of the various terms of the expression above, it is convenient to begin by determining some of the terms in the local coordinate system fixed to the earth, in which $\hat{x}_r$ points east, $\hat{y}_r$ points north, and $\hat{z}_r$ points upward. In this system, $\boldsymbol{\omega}_1 = \omega_e \sin\theta\, \hat{y}_r + \omega_e \cos\theta\, \hat{z}_r$, $\boldsymbol{\omega}_2 = -\omega_e \cos\theta\, \hat{z}_r$, and $\boldsymbol{\omega}_1 + \boldsymbol{\omega}_2 = \omega_e \sin\theta\, \hat{y}_r$. In working out an expression for $\frac{d^2 r}{dt^2}$ in coordinate form, we can then transform these expressions to the final $rr$ system precessing at $\boldsymbol{\omega}_2$ with respect to the $r$ system fixed to the earth. For this purpose, let us adopt an $x_{rr}$-axis that lies in the local horizontal plane at an angle $\phi$ measured clockwise from the $x_r$-axis and a corresponding $y_{rr}$-axis at an angle $\phi$ clockwise from the $y_r$-axis, as shown in Fig. 2 of the main text. We then can take $\phi(t) = \omega_e \cos\theta\, t$. Taking into account that $(d\boldsymbol{\omega}_1/dt)_1$ and $(d\boldsymbol{\omega}_2/dt)_2$ are both zero for the pendulum, the (albeit somewhat cumbersome) result for $\frac{d^2 r}{dt^2}$ is

$$\frac{d^2 r}{dt^2} = \left(\frac{d^2 r}{dt^2}\right)_{rr} + [2\omega_e \sin\theta \cos\phi\, (dz/dt) - \omega_e^2 \sin^2\theta \, \cos\phi(\cos\phi\, x + \sin\phi\, y) - \omega_e^2 \sin\theta \cos\theta \sin\phi\, z]\, \hat{x}$$
$$+ [2\omega_e \sin\theta \sin\phi\, (dz/dt) - \omega_e^2 \sin^2\theta \sin\phi(\cos\phi\, x + \sin\phi\, y) + \omega_e^2 \sin\theta \cos\theta \cos\phi\, z]\, \hat{y}$$
$$+ \left[-2\omega_e \sin\theta \left(\cos\phi \left(\frac{dx}{dt}\right) + \sin\phi \left(\frac{dy}{dt}\right)\right) + \omega_e^2 \sin\theta \cos\theta(\sin\phi\, x - \cos\phi\, y) - \omega_e^2 \sin^2\theta\, z\right]\, \hat{z}, \quad (S1)$$

in which the subscript $rr$ should be understood to apply to all of the coordinates and unit vectors on the right. In this expression we have allowed for the small amount of locally vertical motion undergone by the pendulum bob.

Thus the price that we pay for the second transformation, which eliminates to lowest order the effects of the Coriolis force, is a more cumbersome expression for the higher-order terms than we would have had in the system at rest with respect to the earth's surface. Nevertheless, it puts us in a good position to address the question of whether the higher-order terms disturb in any significant way the simplicity of the motion we have found in the final rotating system. In particular, consider the simplest case, that of initially plane oscillation along the $x$-axis of the precessing system. Of main interest are sidewise effective forces, i.e. forces acting along the precessing $y$-axis, that act to deflect the motion. Effective forces acting along the precessing $x$-axis, along with those acting along the z-axis, would only act to disturb in a minor way the periodicity of the oscillation.

In what follows, let us continue to work in the precessing system and to drop the subscript $rr$ for simplicity, it being understood to apply now to all coordinates. In order to focus on the influence of the effective forces due to rotation, let us assume that the y-axis motion then obeys the differential equation

$$\left(\frac{d^2y}{dt^2}\right) + 2\omega_e \sin\theta \sin\phi \left(\frac{dz}{dt}\right) - \omega_e^2 \sin^2\theta \sin\phi(\cos\phi\, x + \sin\phi\, y) + \omega_e^2 \sin\theta \cos\theta \cos\phi\, z \cong -\left(\frac{g}{l}\right)y. \quad (S2)$$

In so doing, we are leaving out terms on the right-hand side of the equation of higher order in the horizontal displacement and its derivatives that are due to the departure of the true force from a simple harmonic force, and that may be comparable to or even larger than the effective-force terms on the left-hand side. These are the terms that give rise to the lowering of the pendulum frequency and to the natural precession of the pendulum mentioned in the main text. However, the resulting anharmonicity in the pendulum motion and the reduction in frequency would only have higher-order effects on the effective-force terms, and we can minimize any natural precession by restricting our attention to any y-motion that is very much smaller than the x-motion that we start off with.

In order to assess the influence of the y-axis terms, it seems simplest to assume simple periodic motion of the bob along the x-axis and to treat the y-axis terms as perturbations in first order. Thus, let us assume that $x(t) \cong \alpha l \cos(\omega_p t)$, where, as in the main text, $\alpha$ is a dimensionless amplitude coefficient, $l$ is the length of the pendulum suspension, and $\omega_p$ is the angular frequency of the pendulum swing, $\omega_p^2 \cong g/l$. The accompanying vertical motion is given by $z(t) \cong l - \sqrt{l^2 - x^2} \cong \left(\frac{l\alpha^2}{2}\right)\cos^2(\omega_p t)$.

Our perturbation approach then will be to solve the differential equation

$$\left(\frac{d^2y}{dt^2}\right) + \left(\frac{g}{l}\right)y - \omega_e^2 \sin^2\theta \sin^2\phi\, y$$
$$= 2\omega_e \omega_p l\alpha^2 \sin\theta \sin\phi \sin(\omega_p t)\cos(\omega_p t) + \omega_e^2 l\alpha \sin^2\theta \sin\phi \cos\phi \cos(\omega_p t)$$
$$-\omega_e^2 (l\alpha^2/2) \sin\theta \cos\theta \cos\phi \cos^2(\omega_p t). \quad (S3)$$

Here we have grouped the three terms linear in y and its second derivative on the left-hand side, and substituted in the y-independent terms remaining on the right-hand side the lowest-order forms for $x(t)$, $z(t)$, and $dz(t)/dt$. We can now solve the resulting homogeneous equation and the inhomogeneous equations for each of the terms on the right-hand side separately and add the results. Each term is anisotropic in the horizontal plane, but the angle of precession $\phi$ changes so slowly in relation to the period of the pendulum that we can regard it as constant in solving for $y(t)$ here.

The solution of the homogeneous equation is simple harmonic oscillation in which the square of the angular frequency of oscillation is lowered by a term of order $\omega_e^2/\omega_p^2 = 1.70 \times 10^{-8}$ (making use of the representative numerical values given in the main text), a correction so small that we can ignore the term in what follows. No deflection of the motion from the x-direction would result if the initial values of y and its first derivative were zero.

With the first term on the right alone, the equation is that of a simple harmonic oscillator driven at an angular frequency of $2\omega_p$ with the resultant $y$-motion given by

$$y(t) = -(l\alpha^2/3)(\omega_e/\omega_p) \sin\theta \sin\phi \sin(2\omega_p t), \qquad (S4)$$

yielding with the $x$-motion a very narrow 2:1 (figure-of-eight) Lissajous figure with no direct influence on the precession observed in the earth-bound system. For the representative numerical values given near the beginning of the article, the width to length ratio of the figure would be $< 1.3 \times 10^{-6}$.

The second term on the right involves a driving force at the same angular frequency as the natural angular frequency of $y$-motion of the pendulum, and no steady-state solution exists. If we have initially $y = dy/dt = 0$, then with the second term alone we have

$$y(t) = (\omega_p t)(l\alpha/2)(\omega_e^2/\omega_p^2) \sin^2\theta \sin\phi \cos\phi \sin(\omega_p t), \qquad (S5)$$

a response at the natural frequency of the pendulum with an amplitude that increases linearly with time. The phase of this response relative to that of the primary $x$-motion of the pendulum is such that it produces no alteration of the precession of the pendulum in the earth-bound system. Instead, it represents a very slowly increasing ellipticity of the pendulum motion in the precessing system, motion that was assumed to be initially linear. In one full period of the pendulum, the ratio of the semi-minor axis to the semi-major axis would become $\pi(\omega_e/\omega_p)^2 \sin^2\theta \sin\phi \cos\phi$, less than or equal to $5.4 \times 10^{-8}$ based on our numerical values. At a latitude of 45°, after one-quarter of a full rotation of the earth, this ratio would be less than or equal to $5.1 \times 10^{-5}$, an ellipticity that would produce a natural-precession rate of less than or equal to $1.9 \times 10^{-4}$ times the precession rate given by $\omega_e \cos\theta$.

Similar to the situation for the first term on the right, the third term involves a driving force at $2\omega_p$, except that here it is combined with a steady offset. The resultant $y$-motion is given by

$$y(t) = (l\alpha^2/12)(\omega_e/\omega_p)^2 \sin\theta \cos\theta \cos\phi \, [\cos(2\omega_p t) - 3], \qquad (S6)$$

and involves no contribution to precession. The phase of the Lissajous figure resulting from this term differs from that of the first term by $\pi/2 \, rad$, yielding a shallow bowl-shaped curve. For the representative numerical values, the width to length ratio of the figure would be $2.1 \times 10^{-11}$.

Thus, in conclusion, once again, the first-order result for the precession rate of the pendulum given by the alternative approach of this article (and others) appears to be highly accurate.